\documentclass{icrc}

\usepackage{times}
\usepackage{mathptm}
\usepackage{graphicx} 

\newcommand{\greeksym}[1]{{\usefont{U}{psy}{m}{n}#1}}

\newcommand{\unu}{\mbox{\greeksym{n}}}
\newcommand{\umu}{\mbox{\greeksym{m}}}
\newcommand{\utau}{\mbox{\greeksym{t}}}
\newcommand{\uth}{\mbox{\greeksym{q}}}

\newcommand{\uepsilon}{\mbox{\greeksym{e}}}

\newcommand{\uDelta}{\mbox{\greeksym{D}}}
\newcommand{\uOmega}{\mbox{\greeksym{W}}}

\newcommand{\etal}{{\it et al.}}
\newcommand{\numu}{$\unu_{\umu}$}

\begin{document}

\title{An Indirect Search for WIMPs with Super-Kamiokande}
\author[1]{A.~Habig (for the Super-K Collaboration)}
\affil[1]{University of Minnesota Duluth}

\correspondence{Alec Habig $<$ahabig@umn.edu$>$}

\firstpage{1}
\pubyear{2001}


\maketitle

\begin{abstract}
  A potential source of high energy neutrinos is the annihilation of
  Weakly Interacting Massive Particles (WIMPs) collecting in
  gravitational potential wells such as the centers of the Earth, the
  Sun, or the Galaxy.  A search for such a WIMP annihilation signal
  using the Super-Kamiokande (Super-K) detector is presented.  Super-K
  observes 1.1 upward through-going muons per day.  These events are
  caused by high energy (typical $E_\nu \sim 100$~GeV) \numu
  interactions in the rock under the detector, and are generally
  consistent with the expected flux from atmospheric neutrinos.  No
  enhancement of the neutrino signal due to WIMP annihilation is seen,
  so upper limits on the possible flux of WIMPS are set.  These limits
  are compared to those from other such indirect searches, and a
  model-independent method is used to compare the Super-K results with
  direct-detection WIMP experiments.
\end{abstract}

\section{Introduction}

There is growing evidence indicating that non-baryonic cold dark matter
constitutes a major component of the universe's total mass, seen only
via its gravitational influences \citep{Spergel97}.  Current models of
the Universe \citep{Turner2000} suggest that it accounts for $(30 \pm 7)
\% $ of the closure density of the universe.

Weakly Interacting Massive Particles (WIMPs) are a promising cold dark
matter candidate \citep{Jungman96}.  WIMPs, stable particles which arise
in extensions of the standard model, undergo only weak-scale
interactions with matter but have masses ranging from tens of GeV to a
few TeV.  If WIMPs exist, their relic abundance (governed by electroweak
scale interactions) is remarkably close to the inferred density of dark
matter in the universe \citep{Jungman96}.  The lightest supersymmetric
particle (LSP) of supersymmetric theories is a theoretically well
developed WIMP candidate \citep{Jungman96}.  If R-parity is conserved,
the LSP is stable and hence should be present in the Universe as a
cosmological relic from the Big Bang.  A likely candidate for this LSP
is the neutralino \citep{Ellis84}.  Current LEP data and cosmological
constraints impose a lower limit of about 51~GeV and an upper limit of
600~GeV on the neutralino mass \citep{Ellis00}.

In this paper we describe an indirect method to search for such relic
WIMPS using the Super--Kamiokande detector to look for a high energy
neutrino signal resulting from WIMP annihilation in the Earth, the Sun,
and the Galactic Center.  This is in contrast to direct-detection
experiments which look for signatures of WIMP interactions with a
nucleus in a low background detector.  However, both direct and indirect
detection experiments probe the coupling of WIMPs to nuclei, allowing a
comparison of our results with those of direct-detection of dark matter
experiments.

\section{Indirect WIMP Searches using Neutrino-Induced Muons}

If WIMPs are indeed the dark matter composing our galactic halo they
will accumulate in the Sun and Earth.  When their orbits pass though a
celestial body the WIMPs have a small but finite probability of
elastically scattering with a nucleus. If the resulting velocity after
such scattering is less than the escape velocity, they become
gravitationally trapped and eventually settle into the core of that
body.

WIMPs which have accumulated in this way annihilate into $\utau$
leptons, b, c and t quarks, gauge bosons, and Higgs bosons.  Over time,
equilibrium is achieved between capture and annihilation, making the
annihilation rate half of the capture rate.  High energy \numu are
produced by the decay of the annihilation products.  The expected
neutrino fluxes from the capture and annihilation of WIMPs in the Sun
and Earth depend upon the composition and escape velocity of the
celestial body, the flux of WIMPS, and the the WIMP-nucleon scattering
cross-section.
There are many calculations of expected neutrino fluxes from WIMP
capture and annihilation in the Sun and
Earth \citep{Press,Freese,Silk,Krauss,Steigman}.

Recently it has also been noticed that if cold dark matter is
present at the Galactic Center it can be accreted by the central black
hole into a dense spike in the density distribution \citep{Gondolo}.
WIMP annihilations in this region could make it a compact source of high
energy neutrinos.

The energetic \numu resulting from WIMP annihilation could be
detected in neutrino detectors.  The mean neutrino energy ranges from
1/3 to 1/2 the mass of the WIMP, and the neutrinos can undergo charged
current interactions with the rock around the detector to produce muons.
Thus, neutrino-induced muons coming from the direction of the Sun, the
Earth and the Galactic Center could provide a signature of non-baryonic
cold dark matter.

\section{WIMP Searches in Super--K}

The Super--Kamiokande (``Super--K'') detector is a 50,000~tonne water
Cherenkov detector, located in the Kamioka-Mozumi mine in Japan with
1000~m rock overburden.  It is divided by a lightproof barrier into an
inner detector with 11,146 inward-facing 50~cm Hamamatsu Photomultiplier
Tubes (PMTs) and an outer detector equipped with 1,885 outward-facing
20~cm Hamamatsu PMTs serving as a cosmic ray veto counter \citep{SK98}.

Interactions of atmospheric \numu in the rock around the detector
produce upward through-going muons in Super--K energetic enough to cross
the entire detector.  The effective target volume extends outward for
many tens of meters into the surrounding rock and increases with the
energy of the incoming neutrino, as the higher energy muons resulting from
these interactions can travel longer distances to reach the detector.
Upward through-going muons are the signature of the highest energy
portion of the atmospheric neutrino spectrum observed by Super--K, with
the calculated parent neutrino energy spectrum peaking at 100~GeV
\citep{SK99}.  The downward going cosmic ray muon rate in Super--K is
3~Hz, making it impossible to distinguish downward-going
neutrino-induced muons from this large background, restricting
neutrino detection to those events coming from below.

Event reconstruction of a muon is performed using the charge and timing
information recorded by the PMT's.  Muons are required to have $\geq
7$~meters measured path length (E$_{\umu} > 1.6$~GeV) in the inner
detector, resulting in an effective area for upward through-going muons
of $\sim 1200~\mbox{m}^2$ with a trigger efficiency of $\sim$100\%.  The
arrival direction and time is reconstructed for each muon, with the
reconstructed direction of the muon an average of $1.4^\circ$ from the
direction of the parent neutrino.  1416 upward through-muon events have
been observed in 1268 live-days from April 1996 to April 2000.  More
details of the data reduction can be found in \citet{SK98}.

The expected background for a WIMP search due to interactions of
atmospheric $\unu$'s in the rock below the detector is evaluated with
Monte Carlo (``MC'') simulations using the Bartol atmospheric ${\unu}$
flux \citep{Agarwal}, the GRV-94 parton distribution function
\citep{GRV-94}, energy loss mechanisms of muons in rock from
\citep{Lipari93}, and a GEANT-based detector simulation.

Analysis of the most recent Super-K atmospheric neutrino
data \citep{SK01} is consistent with ${\unu}_{\umu} \rightarrow
{\unu}_{\utau}$ oscillations with $\sin^2 2{\uth} \simeq 1$ and $\uDelta
\mbox{m}^2 \simeq 2.5 \times 10^{-3} \mbox{eV}^2$.  Therefore for
evaluating our background we suppress the atmospheric muon neutrino flux
using these oscillation parameters.  
For the Sun, there is an additional background of high energy neutrinos
resulting from cosmic ray interactions in the Sun itself, but this is
about 3 orders of magnitude less than the observed atmospheric ${\unu}$
flux and hence can be neglected \citep{Gaisser91}.  An absolute
normalization for the total neutrino flux was obtained by constraining
the total number of MC events to be equal to the observed events after
taking oscillations into account.  In order to compare the expected and
observed distributions of upward through-going muon events with respect
to the Sun and Galactic Center, each MC event was assigned a random time
based on the arrival times of the observed upward through-going muon
events.  This procedure allowed the angle between each MC muon and any
celestial object to be obtained, and thus the distribution of observed
and expected upward muons about the three celestial bodies.

\section {WIMP Analysis}

We searched for a statistically significant excess of observed
neutrino-induced upward through-going muons compared to the MC
background in cones about the potential source with half angles ranging
from 5 to 30 degrees.  Smaller WIMP masses result in a larger angular
spread of the resulting muons, so different cone angles are used to
ensure the capture of $90\%$ of the signal for a wide range of WIMP
masses.  The different cone angles optimize the signal-to-noise ratio
for various potential WIMP masses.

\begin{figure}[t]
\vspace*{2.0mm} 
\includegraphics[width=8.3cm]{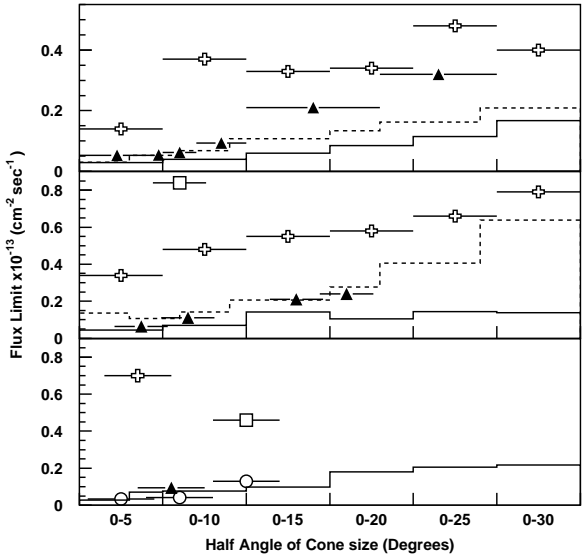}
\caption{Comparison of Super-K excess neutrino-induced upward muon flux 
  limits with those from other experiments for the Earth (top), Sun
  (middle), and Galactic Center (bottom).  Super-K limits are solid
  lines, MACRO dashed lines or circles, Kamiokande crosses, Baksan
  triangles, and IMB squares.  The Y-axis is flux$\times
  10^{-13}cm^{-2}s^{-1}$, the X-axis is half-cone angle size in
  degrees.}
\label{fig-4} 
\end{figure}

No statistically significant excess was seen in any of the half angle
cones.  We can therefore calculate the flux limit of excess
neutrino-induced muons in each of the cones.  The flux limit is given by:
\begin{equation}
\rm \Phi(90\% \, c.l.) = \frac{N_{\cal P}(90\% \, c.l.)}
{\varepsilon (t) \times A(\uOmega)\times T}
\end{equation}
\noindent where $\rm N_{\cal P}$ is the Poissonian upper limit (90\%
c.l.) given the number of measured events and expected background due to
atmospheric neutrinos \citep{PDB} (taking into account oscillations),
and the denominator is the exposure where A($\uOmega$) is the detector
area in the direction of the expected signal ($\uOmega$); $\uepsilon$
is the detector efficiency ($\approx 100\%$ for upward through-going
muons); and T is the experimental livetime.  Figure~\ref{fig-4} shows
the flux limits thus obtained for various cone sizes.

Varying the oscillation parameters applied to the background calculation
only slightly changes the flux limits close to the celestial objects.
The cone with half angle $30^{\circ}$ experiences the largest
fluctuation, with the flux limits varying as much as 10$\%$ for
different oscillation parameters in the neighborhood of the Super--K
allowed region.

The comparison of Super-K flux limits with previous estimates by other
experiments is also shown in Figure~\ref{fig-4}.
All the other experiments have muon energy
thresholds around 1~GeV.  The WIMP flux limits for the Earth and the Sun
by MACRO, Kamiokande, Baksan, and IMB are given 
in \citep{MACRO99,Mori91,Baksan99,IMB86}, and the WIMP flux
limits for the Galactic Center by the above detectors are given in
 \citep{MACRO2,Kamiokande89,Baksan2,IMB87}.
%

%

Once WIMPs are gravitationally captured in the Sun and the Earth they
settle into the core with an isothermal distribution equal to the core
temperature of the Sun or the Earth~\citep{Jungman96}.  Although both
the Sun and the Galactic Center are effectively point sources of
energetic neutrinos resulting from WIMP annihilations, the Earth is not.
Furthermore, muons scatter from the incoming direction of their parent
neutrino due to the kinematics of the initial interaction and multiple
coulomb scattering in the rock below the detector.

\begin{figure}[t]
\vspace*{2.0mm} 
\includegraphics[width=8.3cm]{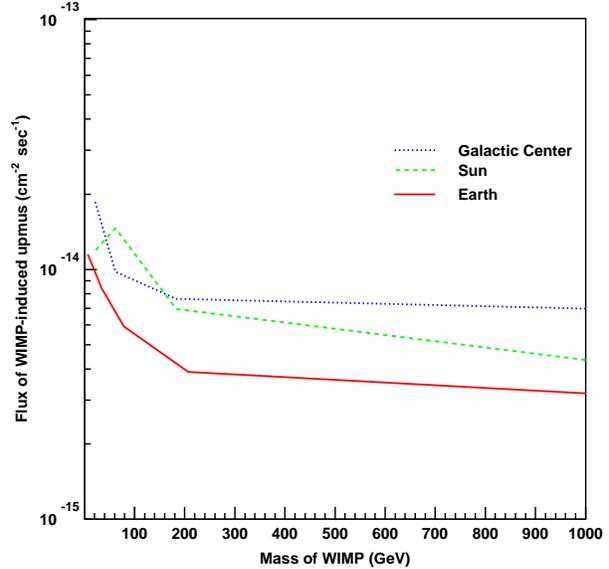}
\caption{Super-K WIMP-induced upward-throughgoing muon flux limits
  from the Earth (solid line); the Sun (dashed line); and the Galactic
  Center (dotted line) as a function of WIMP mass.}
\label{fig-7} 
\end{figure}

The Kamiokande collaboration \citep{Mori91} has calculated the angular
windows for Sun and Earth which contain 90$\%$ of the signal for various
WIMP masses.  
Using these windows, 90$\%$ confidence level flux limits can be
converted to a function of WIMP mass using the cones which collect
90$\%$ of expected signal for any given mass.  These flux limits as a
function of WIMP mass are shown in Figure~\ref{fig-7} for the Earth and
Sun.


\section{Indirect vs. Direct Searches}

In contrast to the indirect WIMP search described above,
direct-detection experiments seek to observe the $\cal{O}$(keV) energy
deposited in a low-background detector when a WIMP elastically scatters
from a nucleus therein.  However, rates for both techniques depend
primarily upon the WIMP nucleon cross-section, either to collide with a
nucleus in the detector, or in a celestial body (and drop below the
excape velocity). The two additional uncertainties which arise in
indirect searches for WIMPs relate to the second moment of the neutrino
energy spectrum and the suppression of annihilation of WIMPs relative to
capture.  Thus, using extreme cases for the above two factors it is
possible to compare the sensitivities of direct and indirect
experiments.  \citet{marc96} calculate the maximum and minimum values of
the ratio of direct to indirect detection rates
for WIMPs with both scalar and axial vector interactions. They find that
the event rate in a 1~kg of Germanium detector is equivalent to that in
$10^{4}-10^{6}$ $m^2$ of upward muon detector.


The DAMA direct-detection experiment reports a detection of WIMPs based
on the annual modulation of their event rate.  Their cumulative analysis
is consistent with the possible presence of WIMP at the 4$\sigma$
confidence level, with the best fit values being : $M_w$ = 52~GeV and
$\sigma_p= 7.2 \times 10^{-6}$~pb \citep{DAMA00}.  The CDMS
direct-detection experiment, however, does not see any WIMP signal.
CDMS data give limits on the spin-independent WIMP-nucleon
elastic-scattering cross-section that exclude parameter space above a
WIMP mass of 10~GeV~c$^{-2}$.  This excludes the entire 3$\sigma$
allowed region for the WIMP signal reported by the DAMA experiment at $>
84\%$~c.l.  \citep{CDMS}.

\begin{figure}[t]
\vspace*{2.0mm} 
\includegraphics[width=8.3cm]{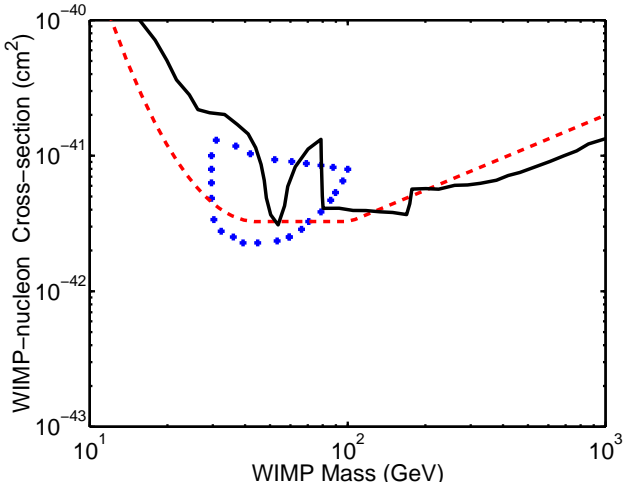}
\caption{Super-K 90\% c.l. exclusion region in WIMP cross section
  vs. WIMP mass parameter space (above solid line), compared to the DAMA
  3$\sigma$ allowed region (inside crosses) and the CDMS 90\% c.l.
  excluded region (above dashed line).}
\label{figsk} 
\end{figure}

Using the results of \citet{marc96} it is possible to obtain limits on
WIMP-nucleon spin-independent cross-section from the Super-K flux limits
and compare them with the results of the DAMA and CDMS direct-detection
experiments.  The combined WIMP flux limits from the Sun and the Earth
as a function of WIMP mass were calculated.  Since the goal is to
calculate an {\it upper limit} on WIMP-nucleon cross-section, 
the most conservative direct/indirect ratio in \citet{marc96} was used
with the Super--K flux limits to calculate the limit on WIMP nucleon
cross-section as a function of WIMP mass.  The Super-K upper limits on
WIMP nucleon cross-section are shown in Figure~\ref{figsk}, along with
the CDMS upper limits and the DAMA best fit region.  These limits rule
out a signification portion of the WIMP parameter space favored by the
DAMA experiment.

\section {Conclusions}

An indirect search for dark matter was performed using 1416
neutrino-induced upward through-going muon events observed by the
Super-K detector corresponding to 1268 days of livetime.  High energy
$\unu$ can be produced from WIMP annihilation in the Sun, the Earth, and
the Galactic Center.  We looked for an excess of upward muons over the
atmospheric neutrino background coming from near those bodies.  No
statistically significant excess was seen.

The lack of excess neutrinos allows flux limits to be obtained for
various cone angles around these potential sources and compared with
previous measurements from other detectors.  For the Sun and the Earth
these flux limits can be calculated as a function of the WIMP mass.  A
comparison of these results with the direct-detection results from CDMS
and DAMA shows a similar sensitivity to a potential WIMP signal over a
wide range of parameter space.

\begin{acknowledgements}
  We gratefully acknowledge the cooperation of the Kamioka Mining and
  Smelting Company. The Super--K experiment was built and has been
  operated with funding from the Japanese Ministry of Education,
  Science, Sports and Culture, and the United States Department of
  Energy. We gratefully acknowledge individual support by the National
  Science Foundation and Research Corporation's Cottrell College Science
  Award.
\end{acknowledgements}

%
%

\end{document}